\documentstyle[art11]{article}
\begin{document}
\title {Basis functions for strongly correlated Fermi  systems}
\vskip 3cm
\author{Mario Salerno\\
\\
salerno@csied.dia.unisa.it
\thanks{Department of Theoretical Physics,
University of Salerno, 84100 Salerno, Italy.}}
\maketitle

\begin{abstract}
A general method to construct basis functions for
fermionic systems which account for the $SU(2)$
symmetry and for the translational invariance of
the Hamiltonian is presented.
The method does not depend on the dimensionality
of the system and it appears as a natural
generalization of the Bethe Ansatz to the case of
non integrable systems. As an
example we present the block diagonalization of
the Hubbard hamiltonian for finite number of
sites in one and two dimensions.
\end{abstract}
\newpage

During the past year a great deal of interest
has been devoted to strongly correlated Fermi
systems in two spatial dimensions because of
their possible role as models for high $T_c$
superconductivity \cite{and87}. Among these
systems, the Hubbard model is certainly
the simplest non trivial model for interacting
electrons in a solid. Its Hamiltonian consists of
a kinetic part, representing the electronic
hopping between next neighbor sites and a
a part representing the repulsive
Coulumbian interaction of electrons with opposite
spins on the same site. In spite of this apparent
simplicity, the mathematical and the physical
properties of this model in dimensions higher
than one (magnetic properties of the ground
state, existence of possible Mott transitions
between insulating and conducting states,
off-diagonal long range order, etc.), are still
poorly understood.  On the other hand, the large
dimension of the Hilbert space of the states
makes any numerical attempt to compute the
spectrum of these systems possible only for
clusters of very small size. A cluster of
$6\times 6$ lattice sites for the Hubbard model
is already behind the computational power (using
algorithms which account for the translational
symmetry and for the conservation of the z
component of the total spin $S$ of the system),
of any supercomputer nowadays available. The main
difficulty is the lacking of general methods
which allow to account for all the symmetries,
both continuous and discrete, of these systems.
For bosonic systems, basis functions which account for
the conservation of the number operator  and for the
translational invariance on a periodic lattice
can be constructed by the so called number state
method \cite{esse92}.
In the case of the 2D Hubbard model one would
like to use as basis functions the simultaneous
eigenfunctions of $S^2$, of the total number
electrons $N$ and of the translational operator
on the lattice. With respect to these functions
the Hamiltonian would acquire a block diagonal
form with blocks of minimal size (if all the
symmetry is included), reducing in a significant
manner the memory needed for the diagonalization
procedure.

The aim of this letter is to present a general method
to construct basis functions for strongly
correlated Fermi systems which span the
irreducible representations (irreps) of $SU(2), U(1)$, and
of the translational group on a periodic lattice.
The method does not depend on the dimensionality
of the system and it appears to be  a natural
generalization of the Bethe Ansatz to the case of
non integrable systems (in the case of the one-dimensional
antiferromagnetic Heisenberg chain it is completely
equivalent to Bethe Ansatz). The method is based
on the following points: i) The invariance of the
total spin of the system under the permutation
group $S_f$ allows to construct basis functions
which span the irreps of $SU(2)$. This is done by
using "filled" Young tableaux for fermionic
systems recently introduced in ref. \cite{msprb95}.
ii) Any discrete group is a subgroup of the permutation
group, thus one can project the above
eigenfunctions on the $S_f$ subgroup of interest
(in the present case on the subgroup
corresponding to the translations on the
lattice). The commutation of the translation
operator with the total spin $S$, assures that the
projected functions will be simultaneous
eigenfunctions of both operators. We illustrate
the method by taking as working example the
Hubbard model both in one and in two dimensions.

The Hubbard Hamiltonian is written as
\begin{equation}
\label{hub}H=-t\sum_{\sigma
{<i,j>}}^fc_{i_\sigma }^{\dagger }c_{j_\sigma }\;{
+\;}U\sum_i^fn_{i_{\uparrow }}n_{i_{\downarrow }}.
\end{equation}
As usual, U represents the onsite Coulomb
repulsion energy between electrons,
$n_{i_{\uparrow }}$ and $n_{i_{\downarrow }}$ are
respectively, spin up and spin down occupation
number operators at site i, t denotes the hopping
energy between sites, $\sigma $ denotes the
possible spin states of the electrons (i.e.
$\sigma =\uparrow $ or $\downarrow ),$ and
$c_{i_\sigma }^{\dagger },c_{j_\sigma }$ are
fermionic creation and annihilation operators
\begin{equation}
\label{ar}\left\{ c_{i_\sigma },
c_{j_{\sigma `}}\right\} =\left\{ c\dagger
_{i_\sigma },c_{j_{\sigma `}}^{\dagger }\right\}
=0,\quad \left\{ c_{i_\sigma },c_{j_{\sigma
`}}^{\dagger }\right\} =\delta _{i,j}\delta
_{\sigma ,\sigma `}.
\end{equation}
As well known the Hamiltonian (\ref{hub})
posses, besides the translational symmetry
and the SU(2) invariance
under rotations of the total spin
$\stackrel{\rightarrow }{S}$ of the system, also the $U(1)$
conservation of the total
number of electrons
\begin{equation}
N=\sum_j\left( n_{j_{\uparrow }}+n_{j_{\downarrow
}}\right),
\end{equation}
and the so called particle-hole symmetry.
The conservation of $N$ and $S_z$ obviously
implies the separate
conservation of the number of electrons with spin
up $ N_{\uparrow }=\sum_jn_{j_{\uparrow }}$ and
of electrons with spin down
$N_{\downarrow}=\sum_j n_{j_{\downarrow }}$. To define the
Hilbert space of the states we denote with $|3>,
|2>,|1>,|0>$ the possible states on a generic
site i.e., respectively, the doubly occupied
state, the single occupied state with spin up,
the single occupied state with spin down and the
vacuum (the numbers 3,2,1,0 are here used as
quantum numbers to characterize them). Since
there are four possible states at each site, the
dimension of the Hilbert space for a lattice with
$f$ sites is just $4^{f}$.  On the other hand the
conservation of the number operator $N$ allows to
decompose the Hilbert space $K$ into a direct sum
of eigenspaces $K_N$ corresponding to a fixed
value of $N$. The dimension of these spaces is
just the number of ways $N$ electrons can be
placed in $2f$ boxes i.e.
$\label{perm}d_N = \frac{(2f)!}{(2f-N)!N!}$.
This leads to the diagonalization (for each $N$)
of a finite $d_N \times d_N$ matrix (we can
restrict $N$ to $N\leq f$ since $f<
N\leq 2f$ follows from the particle-hole
symmetry). We then use the invariance of $S^2$
under $S_f$ to construct $S^2, N$ eigenfunctions
which allow to further block diagonalize each
$d_N\times d_N$ matrix with respect to the
irreps of $SU(2)$. To this end we recall that the
irreducible representations of $S_f$ can be
labelled by all possible partitions $
[f_{1}...f_{k}]$ of $f$ into $k$ parts, with
$f_{i}$ integers obeying $ f_{1}\geq
f_{2}\geq... \geq f_{k}$ and $f_{1}+f_{2}+ ... +
f_{k}=f$. To each partition it is associated a
Young tableau of type $\{f_{1},f_{2},...\}$ i.e.\
having $f_1$ boxes in the first row, $f_2$ boxes
in the second row, etc., each tableau
corresponding to different symmetry class
operations of $ S_f$ \cite{ha62}.

In order to obtain the highest weight vectors of
$SU(2)$ (i.e. eigenvectors of $S^2$, $S_z$
belonging to $S=S_z$) with a given $S_f$ symmetry
we use the idea of Young tableaux ''filled'' with
quanta, introduced for bosonic systems in
ref. \cite{se93} and extended to the
fermionic case in ref. \cite{msprb95}.
To this end we observe that for a
fixed $N=N_{\uparrow }+N_{\downarrow }$, the
possible values of $N_{\uparrow },N_{\downarrow
}$ compatible with it are all the partitions of N
into two parts, each partition being associated
with a well defined value of $S_z$. Let us
introduce the quantum number
\begin{equation}
\label{m}M=3N_{\downarrow }+2(N_{\uparrow }-N_{\downarrow })
\end{equation}
and consider all the partitions
$(m_1,m_2,...,m_f)$ of $M$ into $f$ parts with
$m_i=0,1,2,3$ (in eq. (\ref{m}) and in the
following, we restrict to $ N_{\uparrow }\geq
N_{\downarrow }$ since the cases $N_{\uparrow
}<N_{\downarrow }$ follows from these by
interchanging spin up with spin down electrons).
Since the order of the quantum numbers $m_i$ in
the partition is unimportant we fix it to be
$m_1\geq m_2\geq ...\geq m_f$. We remark that
each partition $(m_1,m_2,...,m_f)$ of $M$ is
associated with an eigenstate $|m_1,m_2,...,m_f>$
of $S_z$. This leads to a family of states
organized into levels. From each $M$ level we
construct eigenmanifolds of $S^2$ with a given
$S_f$ symmetry by filling the quanta $ m_i$,
characterizing that level in the boxes of a Young
tableau according to the following rules: i) The
quanta must be not increasing when moving from
left to right in each row or when moving down
each column of a given tableau. ii) The quanta
referring to spin up and spin down states (i.e.
$m_i=1,2$) cannot appear more than once in a row.
iii) The quanta referring to doubly occupied
states or to empty states (i.e. $m_i=3,0$ )
cannot appear more than once in a column. These
rules directly follow from the permutational
properties of the four states $|3>,|2>,|1>,|0>$
and from the symmetrization and
antisymmetrization property of, respectively,
rows and columns of a given Young tableau.  By
using these rules we construct, for each $M$
level, a family of filled Young tableaux for each
$S_f$ symmetry. To pass from filled tableaux to
states we apply Young symmetrizer and
antisymmetrizer operators which take into account
the Pauli exclusion principle, i.e. every time
the symmetry operator involves the permutation of
$n_a$ spin up electrons and $n_b$ spin down
electrons an extra $(-1)^{n_a+n_b}$ factor is
included (note that due to the permutational
properties of the $|2>$ and $|1>$ states, one
must count only the interchanges of spin
electrons and spin down electrons separately
involved in each permutation). These states, by
construction, are eigenstates of $S_z$ with a
given $S_f$ symmetry but, in general, they are
not eigenstates of $S^2$. In order to identify
the $M$ levels with the eigenmanifolds of $S^2$
we must characterize the filled tableaux
corresponding to highest vectors of $SU(2)$
(highest weight filled tableaux). This
can be done by noting that a change of a 1 into a
2 (1-2 flip) in a filled tableau corresponds to
increases $M$ by $1$ i.e. to pass to a filled
tableau of the $M+1$ level. We have therefore,
that the filled tableaux which survive 1-2 flips
(i.e. the one that satisfy the filling rules also after a
1-2 flip) are the ones for which $S>S_z$. In this way one
''extract'' from each M level the
highest weight vectors of SU(2). We remark,
however, that by a $1-2$ flip two different $M$
tableaux may be associated with the same $M+1$
tableau. In this case one can prove that the
linear combinations $\phi _{\pm }=\psi _1\pm \psi
_2$ of the states corresponding to the $M$ filled
tableaux, produce one $S=S_z$ state ($\phi _{+}$)
and one $S=S_z+1$ state ($\phi _{-}$).
\noindent Note that these states are not necessarily
orthogonal so that, in general, a final
Gram-Schmidt orthonormalized procedure must be
applied. We remark that the functions so
constructed are good basis functions to solve
fermionic systems with infinite-range
interactions (in the thermodynamic
limit this should correspond to exact mean-field
calculations). The application of these functions to
the Hubbard model with unconstrained  hopping is discussed
in ref. \cite{msprb95}. By using these functions one can
easily get the following  characterization of the ground
state for the $S_f$-invariant  Hubbard system for $t>0$:
the N=1,2 ground state is always associated
with a tableau of type $\{f\}$ while for $f\geq
3$ it is associated with tableau of type
$\{f-(N-2),2,...,2\}$ for $N$ even or of type
$\{f-(N-2),2,...,2,1\}$ for $N$ odd.
Furthermore, in the ground state S has always its
minimal value i.e. $S=0$ for $N$ even or $S=\frac
12$ for $N$ odd.

We now come to the problem of projecting the
above functions on the subgroup, say $G$, of $S_f$ of
physical interest (for translations on a
1D periodic lattices $G$ is
just the abelian group $C_f$ corresponding to the
cyclic permutations). This problem is similar to the
one encountered in perturbation theory when a
perturbation reducing the symmetry induces a
splitting in the energy levels. Let us denote by
$ D(R),R\in S_f$ the irreps of $S_f$. A
representation of $G$ is readily obtained by
selecting among the matrices $D(R)$ those
corresponding to elements of $G$. These
representations however are in general reducible
i.e. they can be expressed in terms of irreps
$D^{(\nu )}$ of $ G$ as $D(R)=\sum_\nu c_\nu
D^{(\nu )}(R)$ with $c_\nu $ non negative
integers counting the number of times $D^{(\nu
)}$ appears in $D$.  By denoting with $g_i$ the
number of elements in the class $K_i$ of $G$
and with $g$ the order of this group, one
easily express the integers $ c_\nu $ in terms
of the characters $\chi ,\chi ^{(\nu )}$ of
respectively $ S_f$ and $G$ as
\begin{equation}
\label{splitt}c_\nu =\frac 1g\sum_ig_i\chi _i^{(\nu )\,*}\chi _i.
\end{equation}
This gives the splitting of the irreps of $S_f$
(i.e. of the Young tableaux of a given type) into the
irreps of $G$ (star in Eq.(\ref{splitt}) denotes complex
conjugation). The eigenfunctions $\psi$ of
$N, S^2$, corresponding to the above highest weight
filled Young tableaux, are projected on the $\nu
-$th irrep of $ G$ by using the projection
operator $P^{(\nu )}$ defined by
\begin{equation}
\label{project}\psi ^{(\nu )}=
P^{(\nu )}\psi \equiv \frac{n_\nu }g\sum_R\chi
^{(\nu )\,*}(R)\,\;U_R\cdot \psi
\end{equation}
where the sum is over all the elements
$R$ of the $S_f$ subgroup,  $n_\nu $ is the
dimension of the $\nu$-th irrep of $G$,
$\chi^{(\nu )}(R)$ the corresponding
characters and $U_{R}$ the operator associated to
the group element $R$ \cite{ha62}.
By taking $G$ to be the subgroup
corresponding to the lattice translations $T_n$ and by
projecting all the functions corresponding to
highest weight filled Young tableaux with a
given value of $S$ and $N$,
we get the simultaneous eigenfunctions of
$N,S^2,T_n$ with respect to which the Hamiltonian
aquires block diagonal form.

\noindent Let us illustrate the
method with an explicit calculation on the Hubbard
model with $f=4$. We consider the four sites disposed
in two different configurations:
the first corresponding to a 1-D periodic chain,
the second to a 2-D periodic square lattice.
In the first case the $S_f$
subgroup of interest is the cyclic group $C_4,$
while in the second case is the group $C_{2h}$.
They are both abelian groups with one dimensional
irreps.  Let us denote with
$A, B, E_1, E_2,$ the irrep of $C_4$ and with
$A_u, A_g, B_u, B_g$ the irrep of $C_{2h}$ (we refer to the
standard notation of point-symmetry groups).
The two irrep $E_1,E_2$
of the group $C_4$ are one the complex conjugate
of the other thus they physically correspond to a double
degenerate level (this is true also for other
$C_f$ groups). This accidental degeneracy is
connected to time reversal invariance of the Hubbard
Hamiltonian i.e. the complex conjugate of an eigenfunction is
automatically an eigenfunction with the same
energy. We take advantage of this fact by
considering $E_1,E_2$ equivalent to a single
irrep $E$ of dimension two. For brevity we
concentrate only on the case $S=0$ at half filling
($N=4)$ (a detailed analysis will be published elsewhere
\cite{msjpa95}). In Table 1 we report all the
$S=0$ highest weight filled tableaux together with
their splittings in terms of the irreps of the groups
$C_4$, and $C_{2h}$. From this table we see that for
the 1-D chain one gets two blocks of dimension
$6\times 6$ associated with the irrep $A$ and $B$
(this giving 12 $S=0$ nondegenerate eigenvalues),
and one $4\times 4$ block associated to the irrep
$E$ (giving four doubly degenerate eigenvalues).
In the case of the 2D chain we see that the
accidental degeneracy in the $E$ representation
is removed, and we have three $4\times 4$ blocks
(respectively associated to the irreps
$A_u,B_u,B_g$) and one
$8\times 8$ block associated to the $A_g$
representation. Let us concentrate here
only on the ground state. To this end we remark that for
the $S_f$-invariant Hubbard system the ground state
is characterized by a $S=0$ Young Tableau of type
$\{2,2\}$ (see above discussion and ref. \cite{msprb95} for
details). We conjecture that the projection on the
translational subgroup will not alter this situation, i.e. that
the ground state belongs to one of the irrep in
which the $S_f$ ground state splits (we think this
conjecture holds true for general cases if $f$ is even and $t>0$).
This is indeed what happens for the present cases. We find
that the ground state is of type $B$ for the 1-D
chain and of type $A_g$ for the 2-D chain. The block
associated to the $B$ representation of the 1D chain is given by
\begin{equation}
\left(\matrix{ 2\,U & {{-4\,t}\over {{\sqrt{3}}}} &
-\left( {\sqrt{{8\over 3}}}\,t \right) & 0 & 0 &
0 \cr {{-4\,t}\over {{\sqrt{3}}}} & {{8\,t}\over
3} + U & {{-\left( {2^{{3\over 2}}}\,t \right)
}\over 3} & 0 & 0 & 0 \cr -\left( {\sqrt{{8\over
3}}}\,t \right) & {{-\left( {2^{{3\over 2}}}\,t
\right) }\over 3} & {{-8\,t}\over 3} + U & 0
& 0 & 0 \cr 0 & 0 & 0 & 2\,U & 0 & -2\,t \cr 0 &
0 & 0 & 0 & 0 & 2\,{\sqrt{3}}\,t \cr 0 & 0 & 0 &
-2\,t & 2\,{\sqrt{3}}\,t & U \cr }\right)
\label{1DGS},
\end{equation}
while the block associated to the $A_g$ representation
of the 2D chain is:
\begin{equation}
\left(\matrix{ 2\,U & {{-16\,t}\over 3} & 0 & 0 &
{\sqrt{{{32}\over 3}}}\,t & 0 & 0
& {{{2^{{5\over 2}}}\,t}\over 3} \cr
{{-16\,t}\over 3} & U & {{-\left( {2^{{3\over
2}}}\,t \right) }\over 3} & {\sqrt{{8\over
3}}}\,t & 0 & {\sqrt{{8\over 3}}}\,t & -\left(
{2^{{3\over 2}}}\,t \right) & 0 \cr 0 & {{-\left(
{2^{{3\over 2}}}\,t \right) }\over 3} & 2\,U & 0
& {{4\,t}\over {{\sqrt{3}}}} & 0 & 0 &
{{-8\,t}\over 3} \cr 0 & {\sqrt{{8\over 3}}}\,t &
0 & 0 & -4\,t & 0 & 0 & {{8\,t}\over
{{\sqrt{3}}}} \cr {\sqrt{{{32}\over 3}}}\,t & 0 &
{{4\,t}\over {{\sqrt{3}}}} & -4\,t & U & 0 & 0 &
0 \cr 0 & {\sqrt{{8\over 3}}}\,t & 0 & 0 & 0 &
2\,U & 0 & {{-4\,t}\over {{\sqrt{3}}}} \cr 0 &
-\left( {2^{{3\over 2}}}\,t \right) & 0 & 0 & 0 &
0 & 0 & 4\,t \cr {{{2^{{5\over 2}}}\,t}\over 3} &
0 & {{-8\,t}\over 3} & {{8\,t}\over {{\sqrt{3}}}}
& 0 & {{-4\,t}\over {{\sqrt{3}}}} & 4\,t & U \cr
} \right).
\label{2DGS}
\end{equation}

To check these results we have numerically
diagonalized $H$ in both cases
in the $N=4$ eigenspace of dimension $d_4=70$.
{}From these calculations it follows that,
for $t=1,U=2$ and for $f=N=4$, the ground state
of the 1D chain is non degenerate with energy
$E=-2.82843$ while for the 2D chain the ground
state is non degenerate with energy
$E=-6.681695$. One easily verify that these
values coincide with those obtained by diagonalizing the
blocks respectively in Eq.(\ref{1DGS}) and Eq.(\ref{2DGS}).

In closing this letter we remark that the above method
of block diagonalizing $H$ is quite general and can be applied
to more complicate fermionic systems such as the  Anderson model
as well as to other  subgroups (besides translations) of physical interest
such as the invariance group of fullerene molecules. Furthermore we note that
the study of the Heisenberg model and of the t-J model in one and two
dimensions directly follow from the above analysis
by restricting the space of the single site states respectively
to $|0>,|1>$ (Heisenberg) or  $|0>,|1>,|2>$ (t-J).
We also remark that the method is completely algebraic and
can be easily implemented on a computer
(by using Mathematica \cite{wo91} we have
set up packages which performs all the operations to construct
the above basis functions). We hope our method will facilitate
future numerical studies
of strongly correlated Fermi systems contributing to the understanding
of the physical properties of these systems in two and three dimensions.

\vskip 1.5cm
\noindent{\bf{Acknowledgments}}
\vskip .5cm
\noindent Financial support from the INFM
(Istituto Nazionale di Fisica della Materia),
sezione di Salerno, is acknowledged.
\newpage

\noindent{\bf{Table Captions}}
\vskip 1cm
\noindent Table 1
\vskip .5cm
\noindent Decomposition of the filled Young tableaux
corresponding to $S=0$ highest weight
vectors of $SU(2)$ for $f=N=4$, in terms of the irrep of
the groups $C_4$, and $C_{2h}$. The sum of tableaux
denotes the (plus) linear combination of the
corresponding states.

\newpage
$$
\begin{picture}(400,430)
\put(20,0){\line(200,0){380}}
\put(20,70){\line(200,0){380}}
\put(20,205){\line(200,0){380}}
\put(20,340){\line(200,0){380}}
\put(20,400){\line(200,0){380}}
\put(20,0){\line(0,400){430}}
\put(200,0){\line(0,400){430}}
\put(300,0){\line(0,400){430}}
\put(400,0){\line(0,400){430}}
\put(100,420){$S_4$}\put(250,420){$C_4$}\put(350,420){$C_{2h}$}
\put(60,375){\framebox(13,13){3}}\put(75,375){\framebox(13,13){3}}
\put(90,375){\framebox(13,13){0}}\put(105,375){\framebox(13,13){0}}
\put(250,375){$A$}\put(350,375){$A_g$}
\put(60,350){\framebox(13,13){3}}\put(75,350){\framebox(13,13){2}}
\put(90,350){\framebox(13,13){1}}\put(105,350){\framebox(13,13){0}}
\put(250,350){$A$}\put(350,350){$A_g$}
\put(60,310){\framebox(13,13){3}}\put(75,310){\framebox(13,13){3}}
\put(90,310){\framebox(13,13){0}}\put(60,295){\framebox(13,13){0}}
\put(240,302.5){$B$\,,\,$E$}\put(330,302.5){$A_u$\,,$B_u$\,,$B_g$}
\put(60,270){\framebox(13,13){3}}\put(75,270){\framebox(13,13){2}}
\put(90,270){\framebox(13,13){1}}\put(60,255){\framebox(13,13){0}}
\put(240,262.5){$B$\,,\,$E$}\put(330,262.5){$A_u$\,,$B_u$\,,$B_g$}
\put(42.5,225){\Huge{(}}
\put(60,230){\framebox(13,13){3}}\put(75,230){\framebox(13,13){2}}
\put(90,230){\framebox(13,13){0}}\put(60,215){\framebox(13,13){1}}
\put(110,227){+}
\put(125,230){\framebox(13,13){3}}\put(140,230){\framebox(13,13){1}}
\put(155,230){\framebox(13,13){0}}\put(125,215){\framebox(13,13){2}}
\put(177.5,225){\Huge{)}}
\put(240,222.5){$B$\,,\,$E$}\put(330,222.5){$A_u$\,,$B_u$\,,$B_g$}
\put(60,180){\framebox(13,13){3}}\put(75,180){\framebox(13,13){3}}
\put(60,165){\framebox(13,13){0}}\put(75,165){\framebox(13,13){0}}
\put(240,177.5){$A$\,,\,$B$}\put(340,177.5){$A_g$\,,$A_g$}
\put(60,140){\framebox(13,13){2}}\put(75,140){\framebox(13,13){1}}
\put(60,125){\framebox(13,13){2}}\put(75,125){\framebox(13,13){1}}
\put(240,132.5){$A$\,,\,$B$}\put(340,132.5){$A_g$\,,$A_g$}
\put(40,95){\Huge{(}}
\put(60,100){\framebox(13,13){3}}\put(75,100){\framebox(13,13){2}}
\put(60,85){\framebox(13,13){1}}\put(75,85){\framebox(13,13){0}}
\put(100,97){+}
\put(125,100){\framebox(13,13){3}}\put(140,100){\framebox(13,13){1}}
\put(125,85){\framebox(13,13){2}}\put(140,85){\framebox(13,13){0}}
\put(165,95){\Huge{)}}
\put(240,92.5){$A$\,,\,$B$}\put(340,92.5){$A_g$\,,$A_g$}
\put(40,30){\Huge{(}}
\put(60,45){\framebox(13,13){3}}\put(75,45){\framebox(13,13){2}}
\put(60,30){\framebox(13,13){1}}\put(60,15){\framebox(13,13){0}}
\put(100,30){+}
\put(125,45){\framebox(13,13){3}}\put(140,45){\framebox(13,13){1}}
\put(125,30){\framebox(13,13){2}}\put(125,15){\framebox(13,13){0}}
\put(165,30){\Huge{)}}
\put(240,30){$B$\,,\,$E$}\put(330,30){$A_u$\,,$B_u$\,,$B_g$}
\end{picture}
$$

\end{document}